\documentclass[preprint,12pt]{elsarticle}

\usepackage{amssymb}

\usepackage[caption=false,font=normalsize,labelfont=sf,textfont=sf]{subfig}

\usepackage{amsmath, amsfonts}
\usepackage{amsthm}
\usepackage{array}
\usepackage{algorithm}
\usepackage{algorithmic}

\newtheorem{proposition}{\textit{Proposition}}
\newtheorem{assumption}{\textit{Assumption}}

\usepackage{booktabs}
\usepackage{multirow}
\usepackage{color}

\usepackage[numbers]{natbib}

\journal{Applied Energy}

\begin{document}

\begin{frontmatter}



  \title{Real-Time Operation Strategy of Virtual Power Plants With Optimal Power Disaggregation Among Heterogeneous Resources}


  \author[inst1]{Qixin Chen}
  \affiliation[inst1]{organization={Department of Electrical Engineering, Tsinghua University},
    city={Beijing},
    postcode={100084},
    country={China}}
  \author[inst1]{Ruike Lyu}
  \author[inst1]{Hongye Guo*}
  \author[inst1]{Xiangbo Su}


  \begin{abstract}
    The virtual power plant (VPP) can aggregate flexible resources on the demand side to provide frequency regulation for the grid, helping address the supply-demand balance challenges. When deploying regulation, the VPP disaggregates the requested power adjustment in real time among its internal heterogeneous resources.
    Achieving optimal power disaggregation in this process is challenging due to the temporal coupling characteristics of the resources, the uncertain regulation signals, and the requirement for fast response. Therefore, existing research relies on heuristic methods, such as proportional disaggregation, and fails to leverage the heterogeneity of multiple resources.
    Here, we propose an optimal operation strategy for VPPs to provide regulation, exploiting the complementary characteristics of heterogeneous resources by prioritizing the use of low-cost resources while considering temporal coupling. To reduce the computational overhead of online deployment, we further propose a fast disaggregation algorithm to eliminate the reliance on optimisation solvers.
    We conducted case studies on the operation of a VPP composed of resources including thermostatically controlled loads and industrial production processes. The results verified the reduced operation cost and increased profit of the VPP under the proposed strategy, with only milliseconds of online computation time. We believe that our work can help better exploit demand-side flexibility.
  \end{abstract}

  \begin{graphicalabstract}
  \includegraphics[width=4.5in]{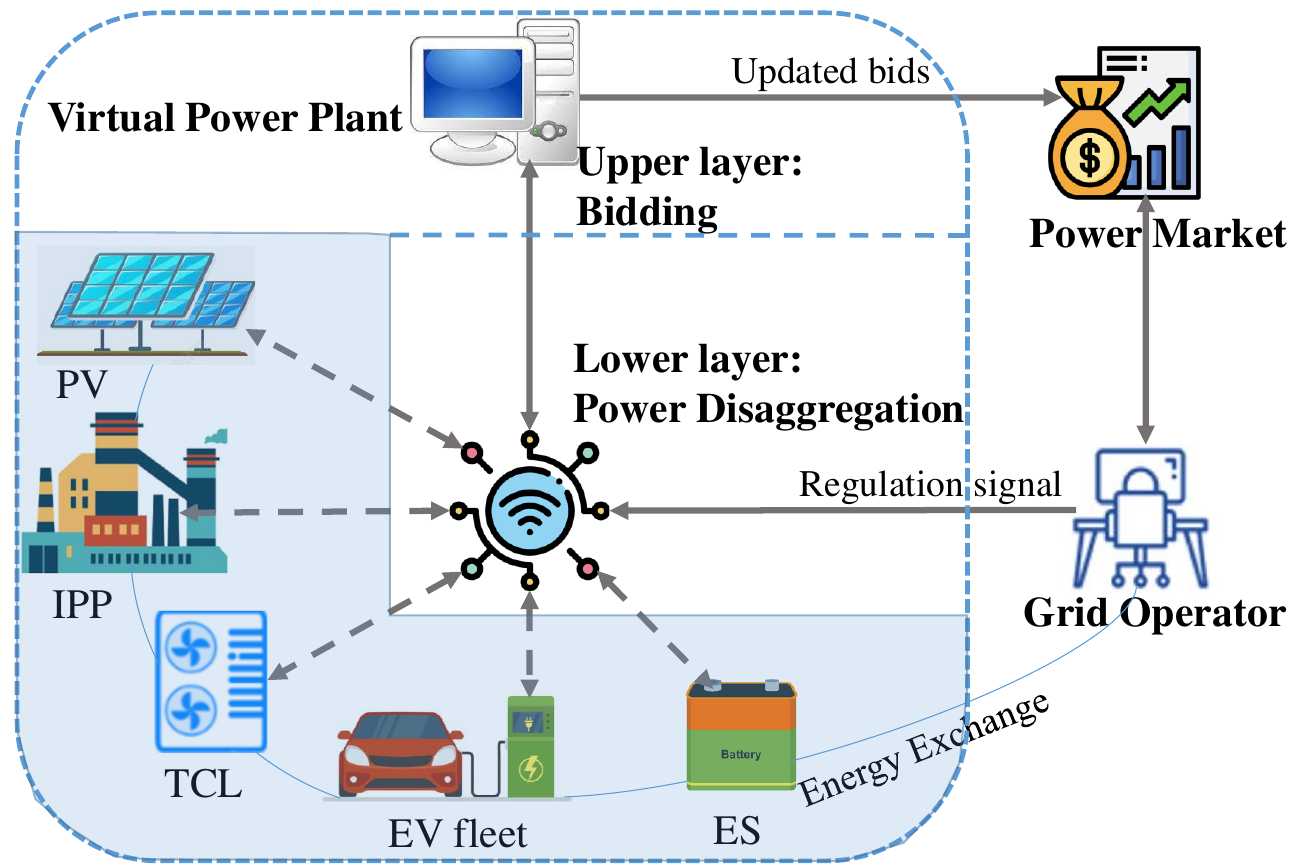}
  \end{graphicalabstract}

  \begin{highlights}
    \item A standardised model of VPP resources, which encompasses industrial production processes as well.
    \item A VPP operational strategy designed to provide regulation services, ensuring optimality over a specific time horizon.
    \item An optimal disaggregation problem that has been scaled down to facilitate online deployment.
    \item A fast disaggregation algorithm that eliminates the need for an optimisation solver.
  \end{highlights}


\end{frontmatter}


\section{Introduction}
Power systems around the world are transitioning away from reliance on fossil fuels. It is estimated that to achieve a 100\% renewable energy power system, wind power and photovoltaics (PVs) in Europe will account for 75\% of the electricity supply~\citep{krakowski_reprint_2016}. This will bring unprecedented challenges to the supply-demand  balance of power systems, as the output of wind power and PVs depends on variable weather, unlike traditional thermal power, which can be flexibly adjusted~\citep{kroposki_achieving_2017}. To address these challenges, a cost-effective approach is to harness the flexibility of demand-side resources, such as thermostatically controlled loads (TCLs), electric vehicles (EVs), and industrial production processes (IPPs). For example, the production process in a cement plant can shift its energy consumption within a day by arranging production schedules, and its power consumption can be quickly adjusted by switching the crushers, providing flexibility at both hourly and minute scales~\citep{zhang_demand_2018}. Nevertheless, due to the relatively small capacity and weak control capabilities of individual demand-side resources, they often need to be aggregated and controlled through a platform called a virtual power plant (VPP)~\citep{wei_bi-level_2018}. Although not physical power plants, VPPs can support the operation of the power system by participating in energy and ancillary service markets, similarly to conventional power plants.~\citep{shayegan-rad_day-ahead_2017}.

Frequency regulation  is a typical ancillary service that a VPP can provide, which helps to cope with grid frequency deviations caused by load fluctuations~\citep{dallanese_optimal_2018}. In providing regulation services under the electricity market environment, a VPP must first determine its regulation bids, that is, the regulation capacity at different hours. Then, upon receiving a regulation signal from the grid operator (GO), the VPP disaggregates the required net power adjustment to its internal resources for response. In practice, resources within a VPP often exhibit heterogeneous operational characteristics, which manifest as parameter differences among resources of the same type and operational constraint differences among resources of different types. Considering such heterogeneity, an ideal power disaggregation strategy should be able to prioritise the use of low-cost resources to respond to the regulation signal. A typical example is that in a VPP composed of battery  storage and wind power, prioritizing the adjustment of wind power output and avoiding the use of batteries with a limited number of charge and discharge cycles can substantially increase the net profit of the VPP~\citep{he_cooperation_2017}.

Nevertheless, in the general case, achieving optimal power disaggregation is much more complex, as the operation of storage-like resources, such as EVs and TCLs, is temporally coupled~\citep{wen_aggregate_2023}, meaning that controlling the output of a certain resource in one period may also affect the profit of the VPP in subsequent periods. For example, frequently reducing the power of the air conditioning of a building at 10 a.m. does not directly generate operating costs, but it may cause the room temperature to be unacceptable at 11 a.m., resulting in the VPP losing the income for regulation services it could have obtained by adjusting the air conditioning power at that time, and may even lead to a need to compensate users for their losses. Theoretically, to optimally respond to a single frequency-regulation signal, the VPP must optimise its operation over the entire horizon (e.g., 24 hours)~\citep{lyu_co-optimizing_2023}. Considering that the time interval of receiving a regulation signal is merely a few seconds and its future value is random,  a large-scale stochastic optimisation problem must be solved online, which could be costly.

Considering the complexity of optimal power disaggregation, existing research on VPP strategies for providing regulation has adopted varying degrees of simplification. The most common simplification is proportional disaggregation, which assumes that the response of each resource to the regulation signal is proportional to the regulation capacity allocated to it in advance. Due to its mathematical simplicity and ease of implementation, proportional disaggregation was initially used to aggregate large-scale EV fleets~\citep{sortomme_optimal_2012} and energy storage (ES)~\citep{vatandoust_risk-averse_2019} and was later widely used in VPPs with different components~\citep{sadeghi_optimal_2021, yi_robust_2022}. However, the strategy of proportional disaggregation does not effectively utilise the heterogeneity between resources to enhance VPP operations. Furthermore, in terms of implementation complexity, the VPP must constantly fine-tune the output of all resources whenever it receives a regulation signal. This can lead to increased communication and control burden, and frequent adjustment in energy usage may considerably inconvenience electricity consumers. Therefore, it becomes challenging for proportional disaggregation to adapt to the requirements of implementing a broader range and more extensive degree of demand response.

Regarding this problem, an intuitive idea is to set response priorities based on the operating characteristics of each resource during power disaggregation  to achieve certain objectives~\citep{hao_aggregate_2015}, such as maximising the welfare of resource owners~\citep{sun_real-time_2014}, ensuring fairness of profits among resources~\citep{escudero-garzas_fair_2012}, and avoiding any impact on energy demand~\citep{vagropoulos_real-time_2016, zhao_geometric_2017}. \citet{peng_optimal_2017} sets the objective of power disaggregation as maximising the regulation capacity that the VPP can provide in the next period, which intuitively can enhance VPP profits. To achieve certain objectives of power disaggregation, solving a small-scale linear programming problem after receiving the regulation signal is typically required~\citep{vagropoulos_real-time_2016}. This process involves repeated online invocation of optimisation solvers, which, although potentially costly, is acceptable in terms of computation time. However, these methods focus on responding to a current regulation signal and fail to achieve optimisation over the entire horizon. Therefore, they are essentially heuristic methods and cannot guarantee long-term optimality.

To summarise, to achieve optimal power disaggregation, it is necessary to comprehensively model the impact of power disaggregation on VPP profits, taking into account the operational differences and temporal-coupling characteristics of various resources, VPP bidding income, and the uncertainty of regulation signals. Moreover, it is crucial to maintain a reasonably low computational complexity for optimal power disaggregation since power disaggregation must be conducted in real time to promptly respond to regulation signals every few seconds. In our previous work~\citep{lyu_co-optimizing_2023}, we proposed a practical framework for EV aggregators to achieve optimal power disaggregation. Specifically, the operation strategy for maximising the aggregator's net profit, which is a large-scale stochastic optimisation problem, was transformed into an equivalent small-scale linear programming problem, thereby facilitating online optimal power disaggregation. Nevertheless, only one type of resource, i.e., EVs, was considered, and online invocation of an optimisation solver was still required. To achieve optimal power disaggregation among different types of resources, standardised modelling can be used to unify the operational costs and constraints of heterogeneous resources, thereby reducing the difficulty of theoretical analysis and control implementation. For example, \citet{yi_aggregate_2021} proposed a standardised model to express the operation of EV, PV, ES, and TCL resources for frequency-regulation services, reducing the difficulty of deriving the aggregate feasible region. Nevertheless, to our knowledge, there is currently no standardised VPP operation model  that can also express the operational characteristics of IPPs. Compared to other resources, the main feature of IPP operation is the process connection between the preceding and succeeding links of the production process, where the intermediate product produced in the preceding stage serves as the raw material for the succeeding stage. As aforementioned, IPPs such as crushers in cement plants can provide considerable flexibility for the grid~\citep{zhang_demand_2018}. In addition, many industrial users are equipped with rooftop PVs, distributed ES and other resources, requiring a standardised model that can simultaneously model IPP and other types of resources to optimise their energy management. Therefore, it would be helpful to have a standardised VPP operation model to provide regulation that integrates the characteristics of IPPs.

In this paper, we extend the optimal operation framework in~\citep{lyu_co-optimizing_2023} to VPPs composed of various distributed resources, such as renewable energy sources (RESs), ES, TCL, and IPP, while eliminating the need to invoke optimisation solvers online. Specifically, we establish a standardised model that can express the operating characteristics of multiple distributed resources, including the pipeline constraints of IPPs. Based on this, we formulate a stochastic optimisation-based optimal operation framework for VPPs providing regulation services that does not rely on simplified assumptions such as proportionality disaggregation. To fit the time step of regulation deployment, we formulate an optimal power disaggregation problem that is equivalent to the original stochastic optimisation over the entire horizon but is far smaller in scale. Further, by utilising the special structure of the optimal power decomposition problem, we propose an algorithm that involves only algebraic operations to obtain a rapid solution, greatly reducing the computational overhead. Case studies verify the increased profits for VPPs providing regulation services under the proposed method and a significant reduction in computational burden.

The main contributions of this paper are summarised as follows:

\begin{itemize}
  \item We propose a standardised model of distributed resources for VPP operation, which can express the operation constraints of typical resources (RES, ES, EV, TCL, and IPP) in a unified form.
  \item We propose a stochastic optimisation-based operational strategy for VPPs to provide regulation services. This strategy prioritises the use of low-cost resources to respond to regulation signals while considering their temporally coupled characteristics, thereby achieving optimality over the horizon. The proposed strategy does not rely on excessive simplification, such as proportion disaggregation or decoupling of bidding and power disaggregation.
  \item To rapidly respond to regulation signals, we formulate an optimal disaggregating problem, which is equivalent to the original stochastic optimisation but with reduced scale, and further propose an algorithm that does not require an optimisation solver. We theoretically prove the optimality of the proposed algorithm and use case studies to verify that the proposed strategy can improve the profit of VPPs in providing regulation services.
\end{itemize}

The remainder of this paper is organised as follows: Section~\ref{sec_framework} presents the overall framework. Section~\ref{sec_model} presents the mathematical formulation of the VPP bidding strategy. Section~\ref{sec_disaggregation} describes the optimal power disaggregation problem and its implementation by a fast algorithm. Section~\ref{sec_case_study} presents a case study with realistic data. Finally, Section~\ref{sec_conclusion} contains the conclusions and prospects for future research.

\section{Overall Framework}\label{sec_framework}

\subsection{Market Framework}

In this paper, we focus on secondary frequency regulation, also known as automatic generation control. Secondary frequency regulation is a process designed to maintain the system frequency close to a nominal value by correcting  frequency and power mismatches that persist after primary regulation. Compared to primary regulation, secondary frequency regulation typically has a slower response, taking place over a period of a few seconds to several minutes, making it suitable for the participation of demand-side resources. Our method can easily be extended to general regulation services, such as spinning reserves. The challenges faced by VPP operation in providing these ancillary services have commonalities, namely, the random nature of their calls and the need for a quick response.

\subsubsection{VPP operation and bidding}
We focus on the optimal operation strategy of a VPP: the compensation or profit allocation to resource owners~\citep{chen_bargaining_2021} is beyond the scope of this paper. We assume that the VPP can conveniently obtain the cost characteristics, operating parameters, and states of its internal resources based on which it formulates the bidding strategy and controls their power consumption/production. Because of the relatively small capacity of the resources, we regard the VPP as a price taker that jointly optimises its bids for energy (baseline power) and regulation capacity in the market, with the forecast market prices as the boundary condition.
Without loss of generality, we use a real-time market framework, in which the VPP can modify bids for future time periods, while the energy and frequency regulation capacity for the current period cannot be modified.

\subsubsection{Real-time regulation service process}
Before actually providing frequency regulation, the VPP  receives its accepted regulation capacity $r$ after market clearing, which should be consistent with its bids since a price taker can declare a regulation capacity with a minimum price to guarantee acceptance.
When actually providing frequency regulation (known as regulation deployment), the GO sends regulation signals $\delta \in [-1, 1]$  to the VPP (e.g., every 2 seconds in PJM~\citep{bjm2022bpm11}). The product of the regulation signal and the regulation capacity $\delta r$ is the to-be-adjusted output of the VPP.

\subsubsection{Uncertainty of the regulation signals}
Regulation signals are generated by an automatic closed-loop feedback control system. From the perspective of the VPP, the parameters of the frequency control system are unknown, and load fluctuations are random. Therefore, a common modelling approach is to divide the possible values of the regulation signal into discrete intervals in which the regulation signal lies~\citep{vagropoulos_optimal_2013}. In this way, the probability of each regulation signal scenario can be predicted based on historical data. We use $s \in [S]$ to denote an arbitrary regulation signal scenario, $[S]$ to denote the set of scenarios, $\delta_s$ to denote the representative signal value, and $\pi_s$ to denote the probability of $s$, respectively. For example, if $s$ represent the scenario where the regulation signal $\delta$ lies in $[0.9, 1)$, then $\delta_s$ could be set to 0.95 and $\pi_s$ can be estimated based on historical data.

\subsection{Framework of the Proposed VPP operation Strategy}

        \begin{figure}[!t]
          \centering
          \includegraphics[width=4.5in]{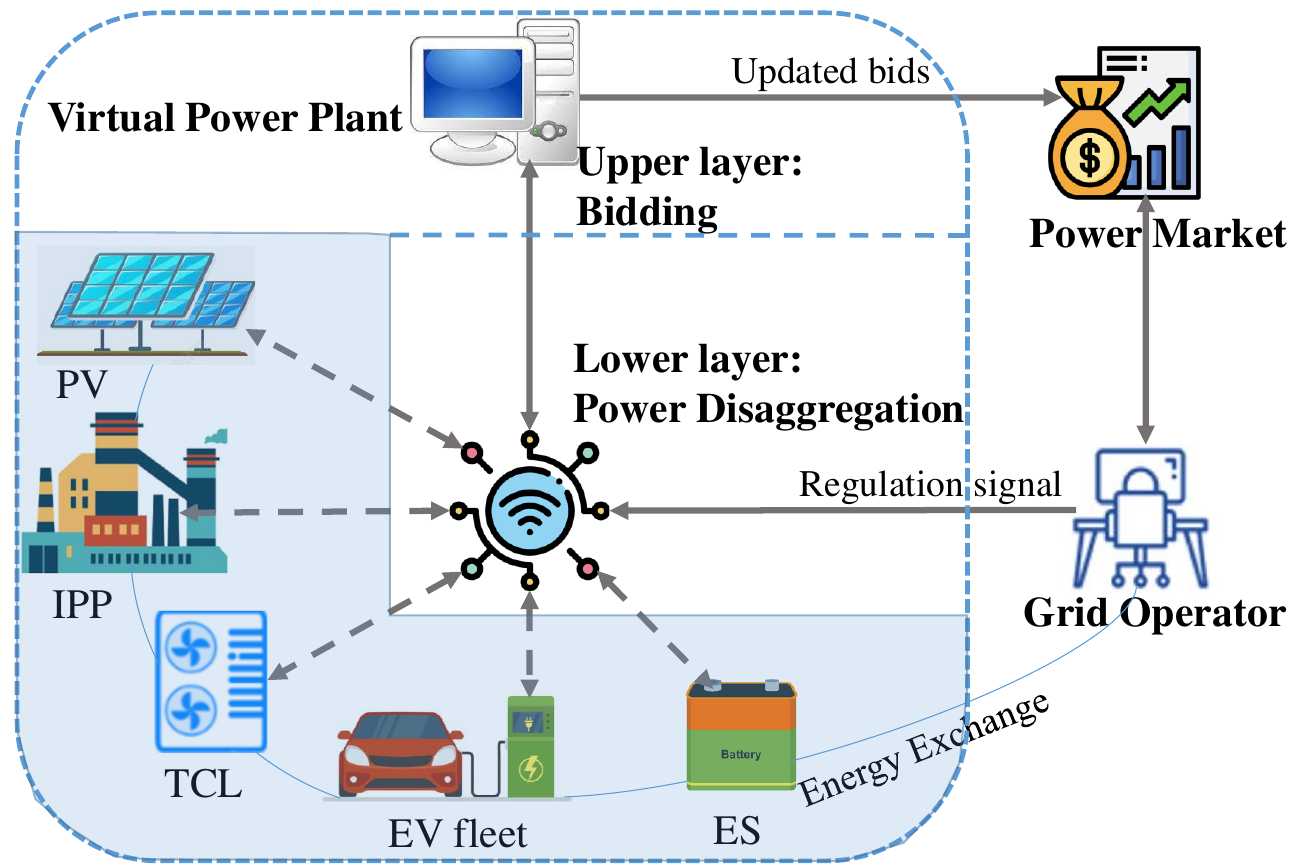}
          \caption{Overview of the proposed framework for VPP operation.}
          \label{fig_overview}
        \end{figure}

        The VPP operation strategy proposed in this paper is divided into an upper layer for bidding strategy and a lower layer for disaggregating power (Fig.~\ref{fig_overview}). At the upper layer, the VPP maximises its bidding profit in the market based on the operational constraints of its internal resources and the boundary conditions of the market. In this process, the VPP must consider the uncertainty of regulation signals and what kind of disaggregation strategy to adopt when deploying frequency regulation and estimate the corresponding operational costs (Section~\ref{sec_model}).

        At the lower layer, the VPP  considers how to control the output of each resource in a fast and optimised manner to respond to the regulation signals sent by the GO. In this process, due to the heterogeneous characteristics of resources and the existence of state variables (such as the state of charge (SOC) of ESs), different control (power disaggregation) strategies not only affect the immediate operating costs produced but may also affect the feasible operating domain of the VPP in future periods, thereby impacting its overall revenue (Section~\ref{sec_disaggregation}).

        \section{Upper Layer: VPP Bidding in the Market}\label{sec_model}

        In this section, we present the mathematical formulation of the VPP bidding strategy. We first introduce the operation models of heterogeneous flexible resources for providing regulation service, based on which we formulate the bidding optimisation problem.

        \subsection{Standardised Operation Model}\label{subsec_standardizedModel}

        In the following content, we use subscripts $t, s, i$ for the relevant variables regarding the time interval $t$, scenario $s$, and resource $i$, respectively. We use subscripts ${\rm ch(dis)}$ and ${\rm init}$ for (dis)charging and initial, respectively. For example, $p^{\rm ch}_{t, s, i}$ represents the charging power of resource $i$ in time interval $t$ in scenario $s$, where "charging" refers to drawing power from the grid. For the denotation of parameters, underlines/overlines are used for lower/upper limits. We use $e$ to denote the state of the resources, which can be the battery SOC, indoor temperature, or products of IPPs. $e_{t, i}$ represents the state of resource $i$ at the beginning of time interval $t$;
        for formal consistency, let $e_{T+1, i}$ denote the state at the end of the time horizon, where $T$ is the number of intervals on the time horizon.

        \subsubsection{Individual Operation Constraints}
        The output $p_{t, s, i}$ of a single resource is the sum of the power in both directions of discharging to the grid and charging from the grid (\ref{model_std_balance2}). This design can consider different characteristics of the two directions while keeping the model linear. Naturally, we have $\underline{p}^{\rm dis(ch)}_{t, i} \ge 0$, and after optimising, the output in both directions will not be positive at the same time. In general, the power limit (\ref{model_std_powerLimit}) and energy limit (\ref{model_std_energyLimit}) of the resources can be time-varying. The expected change in state across time is given by (\ref{model_std_energyChange}), which is affected by the operation of all the associated units. This feature can be observed in IPPs, where operating one process can consume the material of other processes. Also in (\ref{model_std_energyChange}), the dissipation of heat and the effect of ambient temperature on the room temperature is expressed by introducing dissipation rate $\theta$ and $w$, respectively. For a details of the method to obtain the parameter values, refer to ~\citep{liu_optimal_2021}.
        The initial states of the resources are given by (\ref{model_std_energyInit}).
        \begin{subequations}\label{model_std}
          \begin{equation}\label{model_std_balance2}
            p_{t, s, i} =  p^{\rm dis}_{t, s, i} - p^{\rm ch}_{t, s, i}, \forall t, \forall s, \forall i
          \end{equation}
          \begin{equation}\label{model_std_powerLimit}
            \underline{p}^{\rm dis(ch)}_{t, i} \le p^{\rm dis(ch)}_{t, s, i} \le \overline{p}^{\rm dis(ch)}_{t, i}, \ \forall  t, \forall s, \forall i
          \end{equation}
          \begin{equation}\label{model_std_energyLimit}
            \underline{e}_{t, i} \le e_{t, i} \le \overline{e}_{t, i},\ \forall t, \forall i
          \end{equation}
          \begin{align}\label{model_std_energyChange}
            e_{t+1, i} = & \theta_i e_{t, i} +
            \underset{s}{\sum}\pi_{t, s} \underset{j}{\sum} (\eta^{\rm ch}_{ij} p^{\rm ch}_{t, s, j} -  \frac{1}{\eta^{\rm dis}_{ij}} p^{\rm dis}_{t, s, j}) \Delta t \\
            \nonumber    & + w_{t, i} \Delta t \ : \ \lambda^{e}_{t, i}, \ \forall t, \forall i
          \end{align}
          \begin{equation}\label{model_std_energyInit}
            e^{\rm init}_{i} - e_{t, i} = 0,\ t = 1, \forall i
          \end{equation}
        \end{subequations}

        The standardised model encompasses the operating characteristics of all typical resources, and for a specific resource, its operating model can be seen as a degeneration of the standardised model:

        The output of RESs is constrained by the maximum available generation, which can be expressed by (\ref{model_std_powerLimit}). For the operation of ES, the relevant coefficients, such as $\underline{p}^{\rm dis(ch)}_{t, i}$, are actually constants. The EV operation model is similar to that of ES except that the plug-in state indicating whether the EV is connected to the charging pile changes; therefore, relevant coefficients such as $\underline{p}^{\rm dis(ch)}_{t, i}$ become time-varying. The characteristics of deferrable loads are similar to those of EVs, with the primary feature being that the total electricity consumption over a period must reach a value that is acceptable to the user. Due to thermal inertia, the temperature of the TCL can be analogous to the SOC in ES, thereby building a battery model of TCLs~\citep{liu_optimal_2021}. The room temperature requirements may be time-varying, leading to varying $\underline{e}_{t, i}/\overline{e}_{t, i}$; for example, an office building may have a looser temperature restriction during off-hours.

        The modelling of IPP here is based on the linearised state-task model we developed previously~\citep{lyu_lstn_2023}, including the power limit, material buffer limit, initial value, and production target of IPPs.
        The unique characteristic of IPP is that the buffer state of an IPP (i.e., the amount of intermediate goods it produces) is affected not only by its own operating power but also the operation of other IPPs because the product of one IPP may be the raw material of another. This feature can be described by (\ref{model_std_energyChange}), where $\eta^{\rm ch}_{ij}$ is the inter-production efficiency between IPPs, which represents the change in material $i$ caused by the energy consumed by production link $j$. This can be described in the form of an incidence matrix (see Appendix~\ref{app_cptDOFR}), where for other types of resources such as ES, only the diagonal elements are non-zero (i.e., $\eta^{\rm ch}_{ii}$ for the charging efficiency of ES $i$), making it a degenerate case of the model.

        For resources that cannot meet the frequency regulation service requirements, it is assumed that their power is constant in a specific time period and does not participate in the VPP's response to regulation signals. This only requires adding (\ref{model_nc_power}) for the corresponding resource to the bidding model and fixing its output at the reference power when deploying regulation.
        \begin{equation}\label{model_nc_power}
          p^{\rm dis(ch)}_{t, s, i} = p^{\rm dis(ch)}_{t, 0, i}, \ \forall t, \forall s
        \end{equation}

        \subsubsection{Aggregate Power Balance} In each dispatching scenario, the aggregate power of all resources must follow the command sent by the grid (\ref{model_std_balance}).
        \begin{equation}\label{model_std_balance}
          \underset{i}{\sum} p_{t, s, i} = \tilde{p}_t + \delta_s r_t, \forall t, \forall s
        \end{equation}
        where $\tilde{p}_t$ is the energy bid (baseline power) of the VPP in time period $t$ and $r_t$ is the regulation capacity of the VPP in time period $t$. The product of the regulation signal $\delta_s$ and $r_t$ is the to-be-adjusted output of the VPP in scenario $s$.

        \subsubsection{Required maintenance time}
        To ensure the reliability of the regulation capacity, i.e., that the regulation resources can continuously provide deployed regulation under extreme conditions, the GO usually requires that the regulation resources can maintain maximum output for a period of time $\Delta t^{req}$ (e.g., 15 minutes in PJM~\citep{masiello_business_2014}), which can modelled by (\ref{model_std_req1}-\ref{model_std_req2}) $(\forall t, \forall i)$, where the scenarios of $\delta_s = 1$(\ref{model_std_req1}) and $\delta_s = -1$(\ref{model_std_req2}) should be representative. For resources with energy constraints, the longer the required duration, the smaller the available regulation capacity usually is. This essentially increases the conservatism of regulation capacity to cope with the uncertainty of the regulation signal.
        \begin{subequations}\label{model_std_req}
          \begin{equation}\label{model_std_req1}
            (1 - \frac{(1 - \theta_i)\Delta t^{\rm req}}{\Delta t}) e_{t, i} - \underset{j}{\sum}  \frac{1}{\eta^{\rm dis}_{ij}} p^{\rm dis}_{t, s, j} \Delta t^{\rm req} + w_{t, i} \Delta t^{\rm req} \ge \underline{e}_{t, i}, \forall i, \delta_s = 1
          \end{equation}
          \begin{equation}\label{model_std_req2}
            (1 - \frac{(1 - \theta_i)\Delta t^{\rm req}}{\Delta t}) e_{t, i} + \underset{j}{\sum} \eta^{\rm ch}_{ij} p^{\rm ch}_{t, s, j} \Delta t^{\rm req} + w_{t, i} \Delta t^{\rm req} \le \overline{e}_{t, i}, \forall i, \delta_s = - 1
          \end{equation}
        \end{subequations}

        \subsubsection{Operation Cost}
        The cost function of the VPP is given in the  piecewise linear form (\ref{model_std_cost}), where the cost coefficient $Pr$ can also be understood as the compensation price negotiated between the VPP and the resource owners, such as rewarding at 120\% of the cost, rather than the actual cost. If needed, a more accurate piecewise linear model can also be adopted~\citep{han_practical_2014}, without interfering with the linearity of the model.
        \begin{equation}\label{model_std_cost}
          Cost_{t} = \underset{i}{\sum} \underset{s \in [S]}{\sum}\pi_{t, s} (Pr^{\rm dis}_{i} p^{\rm dis}_{t, s, i} + Pr^{\rm ch}_{i} p^{\rm ch}_{t, s, i}) \Delta t, \ \forall t
        \end{equation}

        \subsubsection{Operation Profits}
        Over the time horizon, the VPP's expected profits as a whole in the market are formulated as follows:
        \begin{subequations}\label{model_vpp_profit}
          \begin{equation}\label{model_vpp_profit_total}
            Profit = \sum_{t=1}^{T}(Income^{\rm e}_{t}
            + Income^{\rm r}_{t} - Cost_{t})
          \end{equation}
          \begin{equation}\label{model_vpp_profit_e}
            Income^{\rm e}_{t} =  Pr^{\rm e}_{t} (\tilde{p}_t + \sum_{s} \pi_s  \delta_s r_t) \Delta t
          \end{equation}
          \begin{equation}\label{model_vpp_profit_r}
            Income^{\rm r}_{t} =  s^{\rm perf} r_{t} (Pr^{\rm cap}_{t} + Pr^{\rm mil}_{t} a^{\rm mil}_{t})   \Delta t
          \end{equation}
        \end{subequations}
        where $Pr$ denotes the market price, with superscripts e, r, cap, and mil representing energy, regulation, regulation capacity, and regulation mileage, respectively.
        In the income from the energy market (\ref{model_vpp_profit_e}), $\sum_{s} \pi_s  \delta_s r_t$ is the energy exchanged with the grid following regulation signals, which is settled at the energy price.
        In the income from the regulation market (\ref{model_vpp_profit_r}), $s^{\rm perf}$ is the performance score of the VPP, which is a measure of the VPP's performance in providing regulation service, and $a^{\rm mil}_{t}$ is the expected regulation mileage of the VPP in time period $t$, both regarded as known parameters given by the GO.

        \subsection{Consideration of Disaggregation Strategy in Bidding}
        The VPP must model its disaggregation strategy appropriately during bidding, as different disaggregation strategies may result in different operating costs and available regulation capacities, thereby affecting the optimality of the bidding decision.
        Proportional disaggregation is a commonly used strategy in the existing literature: it pre-allocates regulation capacity at the resource level for each resource and assumes that each resource's response to the regulation signal (i.e., bias to its baseline output $p_{t, i, 0}$) is proportional to the allocated regulation capacity $r_{t, i}$ (\ref{model_ppdis_allocation}). The total regulation capacity of the VPP is the sum of each resource's regulation capacity (\ref{model_ppdis_rch}).
        \begin{subequations}\label{model_ppdis}
          \begin{equation}\label{model_ppdis_allocation}
            p_{t, s, i} = p_{t, 0, i} + r_{t, i} \delta_s, \forall t, \forall s, \forall i
          \end{equation}
          \begin{equation}\label{model_ppdis_rch}
            r_{t} = \underset{i}{\sum} r_{t, i}, \forall t
          \end{equation}
        \end{subequations}

        From an economic perspective, proportional disaggregation fails to prioritise the use of low-cost resources to reduce response costs, which can decrease the profits of the VPP. Moreover, in terms of implementation difficulty, the VPP must adjust the output of all resources each time it receives a regulation signal, which may bring greater communication and control burden, and frequent changes to energy use may greatly bother electricity users. For secondary frequency regulation services, the response time up to minutes is sufficiently long for the operation of demand-side flexible resources, such as ES and EV. In this case, the regulation disaggregation strategy of the VPP can be optimised. If the output of each resource in different frequency regulation signal scenarios (i.e., $p_{t, s, i}$) in the operation model is directly regarded as an independent variable, then the optimal solution of the following bidding problem is equivalent to adopting the optimal disaggregation strategy:
        \begin{subequations}\label{model_bidding_opdis}
          \begin{equation}\label{model_bidding_opdis_objective}
            {\rm max.} \ Profit
          \end{equation}
          \begin{equation}\label{model_bidding_opdis_constraints}
            {\rm s.t.} \ (\ref{model_std} - \ref{model_vpp_profit})
          \end{equation}
        \end{subequations}
        In the bidding model, the decision variables for bids are $\{\tilde{p}_{t}, r_{t} | t \in [T]\}$. The power disaggregation strategy is embodied in the decision variables $\{p_{t, s, i} | t \in [T], \forall i, \forall s\}$. In practice, the bids for time intervals that have passed the gate closure time (e.g., $\tilde{p}_{1}, r_{1}$) should be set to the most recently declared value, considered as problem parameters rather than decision variables. Notably, despite having incorporated the power disaggregation strategy in the bidding model, the VPP still requires a power disaggregation method that can run in real time because we have discretised the values of the regulation signal in the bidding model to estimate the expected operation costs, causing the model to not fully align with reality~\citep{lyu_co-optimizing_2023}.

        The primary focus of this paper is to address optimal power allocation among diverse resources. Thus, we have adopted a general approach that does not delve into uncertainties associated with resource parameters such as EV parameters or owner behaviour. If specific requirements necessitate the consideration of uncertainties, the bidding model and power allocation model can be customised accordingly. From a practical standpoint, in scenarios where the VPP comprises resources like air conditioning in commercial buildings and industrial loads, the accompanying EV resources are typically owned by individuals working in these establishments. Hence, the EVs we consider exhibit fleet-like characteristics, with predetermined arrival and departure times. Therefore, the model neglects uncertainties related to resource parameters, which is a common assumption.~\citep{kaur_coordinated_2019}.
        \section{Lower Layer: Power Disaggregation Optimisation}\label{sec_disaggregation}

        \subsection{Optimal Disaggregation Problem}

        When receiving the regulation signal $\delta_s$ at time $\hat{t}$ within time period $t$, the VPP solves an optimal power disaggregation problem to determine the output of each resource within the effective duration $\Delta \hat{t}$ of the regulation signal, until the next regulation signal is received. In this section, we focus on the current time period ($t=1$) and a specific regulation signal, but the subscripts of $t$ and $s$ are retained for formal consistency. The optimal disaggregation problem is formulated as follows:
        \begin{small}
          \begin{subequations}\label{model_opdis_objective}
            \begin{equation}\label{model_opdis_objective_total}
              {\rm min.} \ \underset{i}{\sum} (Cost_{i}^{\rm imd} + Cost_{i}^{\rm prf})
            \end{equation}
            \begin{equation}\label{model_opdis_objective_immediate}
              {\rm s.t.} \ Cost_{i}^{\rm imd} = (Pr^{\rm dis}_{i} p^{\rm dis}_{t, s, i} + Pr^{\rm ch}_{i} p^{\rm ch}_{t, s, i}) \Delta \hat{t}, \ \forall i
            \end{equation}
            \begin{equation}\label{model_opdis_objective_profit}
              Cost_{i}^{\rm prf} =   \underset{j}{\sum} \lambda^{e}_{t, j} (\eta^{\rm ch}_{ij} p^{\rm ch}_{t, s, i} -  \frac{1}{\eta^{\rm dis}_{ij}} p^{\rm dis}_{t, s, i})\Delta \hat{t}, \ \forall i
            \end{equation}
            \begin{equation}\label{model_opdis_balance}
              \underset{i}{\sum} p_{t, s, i} = \tilde{p}_t + r_t \delta_s
            \end{equation}
            \begin{equation}\label{model_opdis_disaggregation}
              \ p_{t, s, i} =  p^{\rm dis}_{t, s, i} - p^{\rm ch}_{t, s, i}, \ \forall i
            \end{equation}
            \begin{equation}\label{model_opdis_powerLimit}
              \underline{p}^{\rm dis(ch)}_{t, i} \le p^{\rm dis(ch)}_{t, s, i} \le \overline{p}^{\rm dis(ch)}_{t, i}, \ \forall i
            \end{equation}
          \end{subequations}
        \end{small}
        The objective of the VPP is to minimise the cost generated by responding to the regulation signal during $\Delta \hat{t}$ (\ref{model_opdis_objective_total}). This cost includes the immediate operating costs $Cost_{i}^{\rm imd}$ (\ref{model_opdis_objective_immediate}) induced within $\Delta \hat{t}$, as well as the impact on the VPP's total profit over the time horizon $Cost_{i}^{\rm prf}$ (\ref{model_opdis_objective_profit}) after the states of the resource are changed (e.g., ES SOC). In (\ref{model_opdis_objective_profit}), $\lambda^{e}_{t, j}$ is the Lagrange multipliers corresponding to the state constraint (\ref{model_std_energyChange}) for resource $j$ and time $t$ at the optimum of the bidding problem.
        If $Cost_{i}^{\rm prf}$ is ignored, the resulting control strategy will essentially be a myopic greedy algorithm. We will show that by incorporating $Cost_{i}^{\rm prf}$ into the objective function, the control strategy becomes optimal.

        The constraints (\ref{model_opdis_balance}-\ref{model_opdis_powerLimit}) are basically consistent with the constraints in the bidding model regarding $t$, but the constraints related to the state variables (\ref{model_std_energyLimit},\ref{model_std_energyChange}) are omitted because we assume the resources to be flexible (Assumption~\ref{assumption_feasible}), implying that their state constraints usually do not take effect in the middle of the time period. On one hand, considering that the duration of a regulation signal is typically only a few seconds, it is reasonable to assume that the state of each resource will not suddenly violate the constraints after responding to a signal. On the other hand, if the state constraint of a resource during the current period is indeed tight or it must use electricity at a power not lower than a specific value to ensure the demand for the next period, then the parameter value of $\underline{p}^{\rm dis(ch)}_{t, i}/\overline{p}^{\rm dis(ch)}_{t, i}$ in (\ref{model_opdis_powerLimit}) can be modified while retaining the form. For example, if the SOC of an ES reaches the allowed lower limit in advance due to model accuracy issues, its discharge power upper limit $\underline{p}^{\rm dis}_{t, i}$ can be set to 0.

        \subsection{Optimality of the Disaggregation Strategy}
        Here, we analyse the optimality of the above disaggregation strategy (\ref{model_opdis_objective}), meaning that the obtained power disaggregation result is consistent with the result given by the optimal disaggregation strategy in the bidding problem (\ref{model_bidding_opdis}).

        Intuitively, at the optimum of the bidding problem, $\lambda^{e}_{t, i(j)}$ represents the shadow price of state $e_{t, i}$, e.g., the change in VPP operation profit caused by every 1 kWh change in ES SOC. $Cost_{i}^{\rm prf}$ given by (\ref{model_opdis_objective_profit}) as the product of $\lambda^{e}_{t, i}$ and the change in $e_{t, i}$ is exactly equal to the effect of the power disaggregation result on the VPP profit over the time horizon. Therefore, even though the decision variables of the disaggregation strategy (\ref{model_opdis_objective})  include only the output of each resource in the current period, it achieves global optimisation of VPP profit.

        The mathematical proof is based on an assumption on the flexibility of the resources and the following propositions (see the Appendices for the proofs):
        \begin{assumption}\label{assumption_feasible}
          The bidding problem (\ref{model_opdis_objective}) and the power disaggregation problem (\ref{model_bidding_opdis}) are strictly feasible.
        \end{assumption}
        \begin{proposition}\label{theorem_optimality}
          \textbf{Optimality.} $\forall s \in [S]$, the optimal solution to the power disaggregation problem (\ref{model_bidding_opdis}), i.e., $\{p_{t, s, i}, \forall i\}$ corresponds to the optimal solution to the bidding problem.
        \end{proposition}
        \begin{proposition}\label{theorem_invariance}
          \textbf{Time invariance.} If the occurrence frequency of the regulation signals is consistent with the forecasted probability and the VPP controls the output of the resources according to the optimal solution of the bidding problem from $\hat{t}$ to $\hat{t}'$ $(\hat{t}' > \hat{t})$ within the current time period, then the optimal solution to the bidding problem remains unchanged.
        \end{proposition}
        According to Propositions~\ref{theorem_optimality} and~\ref{theorem_invariance}, the proposed optimal disaggregation strategy can optimise the net profit of the VPP over the time horizon, and this optimality is time-invariant. Theoretically, the bidding problem needs to be solved only once before entering the current period to obtain $\lambda^{e}_{t, i}$, and the optimal disaggregation problem can be invoked after receiving a regulation signal. If the occurrence of the regulation signals deviates significantly from the forecast, the bidding problem can be solved again to update $\lambda^{e}_{t, i}$.

        \subsection{A Fast Optimal Disaggregation Algorithm}
        The above optimal disaggregation problem involves constraints and variables of only the current period and the specific regulation signal, so the problem size is much smaller than that of the bidding problem. Nevertheless, its solution as a linear program generally requires the use of a commercial solver. Here, we further analyse the structure of this problem and propose an algorithm that involves only algebraic operations to reduce the computational complexity.

        For the sake of clarity, we first make the following variable substitution. The original power segments are re-indexed with $k$, i.e., $p_k$ for the output of an arbitrary power segment, taking injection into the grid as positive. For a discharge segment, we have $p_k(i) := p^{\rm dis}_{t, s, i}$. For a charging segment, we have $p_k(i) := - p^{\rm ch}_{t, s, i}$. For resources with piecewise linear costs, each segment can be represented separately. For the equivalent cost coefficients of the discharging and charging power segments, the following substitutions are made, respectively:
        \begin{small}
          \begin{subequations}\label{model_substitution}
            \begin{equation}\label{model_substitution_dis}
              c_{k}(i) := Pr^{\rm dis}_{i} - \sum_j \lambda^{e}_{t, j} (\frac{1}{\eta^{\rm dis}_{ij}}),
            \end{equation}
            \begin{equation}\label{model_substitution_ch}
              c_{k}(i) := - Pr^{\rm ch}_{i} - \sum_j \lambda^{e}_{t, j} \eta^{\rm ch}_{ij},
            \end{equation}
          \end{subequations}
        \end{small}
        After substituting $p_k$ and $c_k$, the optimal disaggregation problem (\ref{model_opdis_objective}) can be rewritten as:
        \begin{small}
          \begin{subequations}\label{model_alg}
            \begin{equation}\label{model_alg_objective}
              {\rm min.} \ \sum_k c_{k} p_{k}
            \end{equation}
            \begin{equation}\label{model_alg_balance}
              {\rm s.t.} \ \sum_k p_{k} = p^{\rm req}
            \end{equation}
            \begin{equation}\label{model_alg_powerLimit}
              \underline{p}_{k} \le p_{k} \le \overline{p}_{k}, \ \forall k
            \end{equation}
          \end{subequations}
        \end{small}
        where $p^{\rm req} := \tilde{p} + \delta_s r_t$ is the required net output of the VPP and $\underline{p}_{k} / \overline{p}_{k}$ is the lower/upper limit of  power segment $k$, which can be obtained by substituting $\underline{p}^{\rm dis}_{t, i} / \overline{p}^{\rm dis}_{t, i}$ and $- \overline{p}^{\rm ch}_{t, i} / -\underline{p}^{\rm ch}_{t, i}$, respectively.

        The structure of the problem is now more apparent, namely, its inequality constraints (\ref{model_alg_powerLimit}) and cost functions (\ref{model_alg_objective}) of the power segments are decoupled, with the variables only coupled through the equality constraint (\ref{model_alg_balance}). Therefore, a straightforward solution approach is to start by adjusting the power from the segment with the smallest cost coefficient until the total power reaches $p^{\rm req}$ (Fig. ). Based on the substituted variables, we implement this process in Algorithm~\ref{alg_control}.

        \begin{figure}[!t]
          \centering
          \includegraphics[width=2.5in]{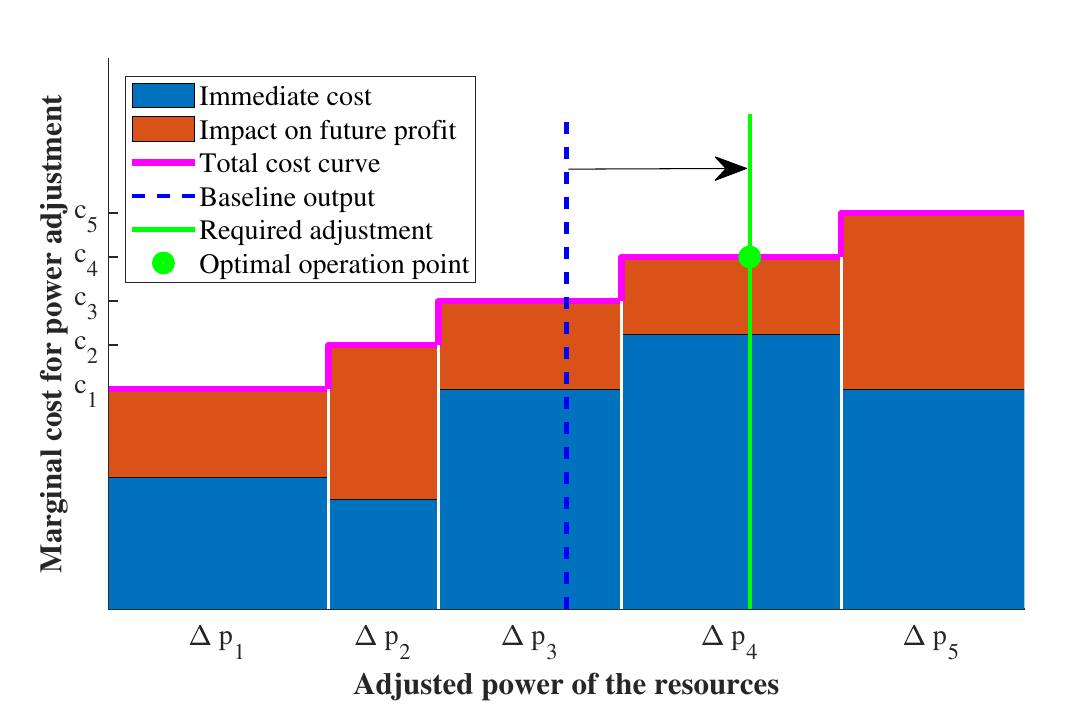}
          \caption{The visual representation of the Fast Disaggregation algorithm. The VPP originally operating at the baseline output point (blue dashed line). In order to achieve the required power adjustment for frequency regulation signals (green line), a two-step approach is employed. Firstly, based on the original resource operating costs (immediately incurred) and the opportunity cost of resource output adjustments derived from the optimal operation problem (influencing future profits), the marginal adjustment costs for each resource are calculated. Subsequently, within the constraints of resource output, the utilisation of low-cost resources is selectively invoked.}
          \label{fig_overview}
        \end{figure}

        \begin{algorithm}[!t]
          \caption{Fast Disaggregation.}
          \label{alg_control}
          \begin{algorithmic}[1]
            \renewcommand{\algorithmicrequire}{\textbf{Input:}}
            \REQUIRE
            required power $p^{\rm req}$, power segment parameters $\{c_k, \underline{p}_k, \overline{p}_k\}$ (indexed in ascending order of $c_k$), number of power segments $K$
            \renewcommand{\algorithmicrequire}{\textbf{Output:}}
            \REQUIRE Updated output of the resources $\{p_{t, s, i}\}$.
            \renewcommand{\algorithmicensure}{\textbf{Initialisation:}}
            \ENSURE $p_k \leftarrow \underline{p}_k, \forall k$, $\Delta p \leftarrow p^{\rm req} - \sum_k p_k$, $k \leftarrow 1$.
            \WHILE{$\Delta p > 0$ and $k \le K$}
            \IF{$\overline{p}_k - p_k < \Delta p$}
            \STATE $p_k \leftarrow \overline{p}_k$, $\Delta p \leftarrow \Delta p - \overline{p}_k + \underline{p}_k$, $k \leftarrow k + 1$.
            \ELSE
            \STATE $p_k \leftarrow \underline{p}_k + \Delta p$, $\Delta p \leftarrow 0$.
            \ENDIF
            \ENDWHILE
            \STATE recover $\{p_{t, s, i}\}$ through $\{p_k\}$ and (\ref{model_opdis_disaggregation}).
          \end{algorithmic}
        \end{algorithm}

        For the input of Algorithm~\ref{alg_control}, the operations involved are parameter substitution based on the optimal solution of the bidding problem and sorting of each resource parameter, which  needs to be calculated only once before entering the current period. After receiving a regulation signal, the complexity of the fast disaggregation algorithm is linear, with only O(5K), where K is the number of power segments. In actual operation, the output of each power segment from the last response can be used as the initial value, rather than re-initializing each time, further reducing the calculation amount. When the rate of change of the frequency modulation signal is not large relative to the length of the power segment of the resources, only the output of the marginal resources needs to be updated, while other resources operate at the upper/lower power boundaries, which is simpler for the implementation of the power control.

        In cases where the actual distribution of the frequency regulation signal deviates significantly from the expected values, resulting in the VPP's inability to fully track the frequency regulation signal, such as when an excessively high upward adjustment is required, Algorithm~\ref{alg_control} adjusts the output of all resources (considering injection into the grid) to their maximum and terminates at $k=K$. This approach aligns with the existing literature, which relaxes the constraint of tracking the frequency regulation signal by introducing a deviation variable and minimises the deviation using the Big-M method~\citep{lyu_co-optimizing_2023}. In practice, due to the limitations imposed on maintenance time (Equation (\ref{model_std_req})), the reported regulation capacity by the VPP exhibits a certain level of conservatism, thereby minimising the probability of such scenarios occurring.

        \section{Case Study}\label{sec_case_study}

        In this section, we demonstrate the effectiveness of the proposed method in enhancing the profits of a VPP for providing regulation services. The VPP consists of PV, ES, a fleet of EVs, TCLs, and IPP resources, with a photovoltaic installation capacity of 2.5 MW and storage capacity of 1 MW/2 MWh. The parameters for the other resources were taken from \citep{chen_scheduling_2021, lu_data-driven_2021, li_emission-concerned_2013}, listed in \citep{lyu_test_data_2023}. We use the system real-time energy and frequency regulation prices of the PJM in July 2022 as boundary conditions (Fig.~\ref{fig_price_bid}(a)). Other market parameters include the length of (scheduling) time interval $\Delta t = 1$ hour,
        the required maintenance time $\Delta t^{\rm req} = 0.5$ hours, and
        the performance score $s^{\rm perf} = 0.984$ given by example 1 provided by PJM in~\citep{example_pjm}. The historical regulation signals (RegD) published by PJM are used for the simulation.

\begin{figure}[!t]
  \centering
  \subfloat[]{
    \includegraphics[width=2.5in]{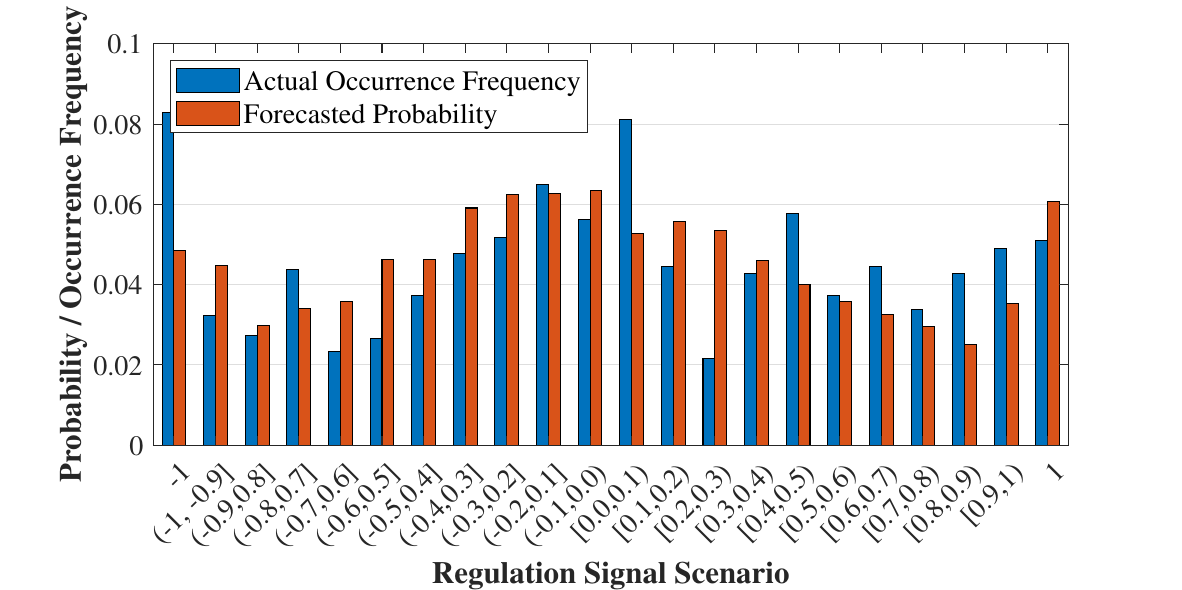}}
  \subfloat[]{
    \includegraphics[width=2.5in]{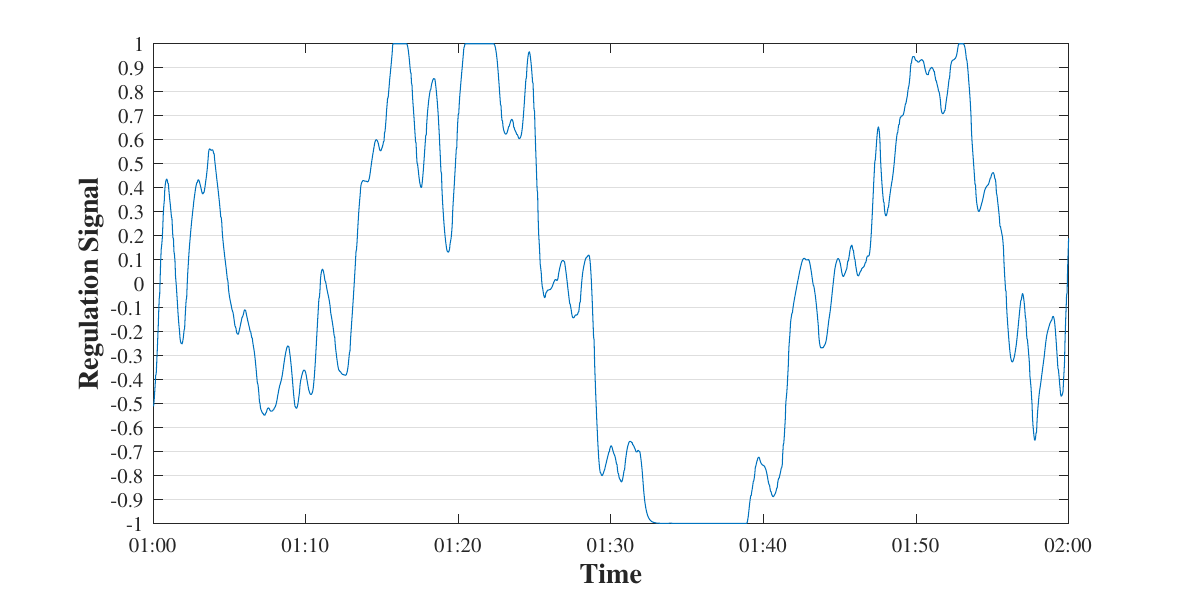}}
  \caption{Regulation signals at 1:00-2:00 on July 21: (a) Actual vs. forecasted distribution of regulation signal scenarios. (b) Actual regulation signals across the hour.}
  \label{fig_signal}
\end{figure}

        We do not assume that the regulation signals can be accurately predicted; it is necessary to estimate the probability of each signal scenario based on historical data. For simplicity and easy reproduction, we divided the nearly continuously distributed frequency regulation values in the range $[-1, 1]$ into 22 scenarios (signal value ranges): $\{-1, (-1, -0.9], …, 1\}$, and used the frequency of the same time period over the past 14 days to estimate the probability of the corresponding scenario. The frequency regulation mileage was estimated in the same way. In the bidding model, the median of the range was used to represent the signal value of the corresponding scenario, for example, $\delta_s= - 0.95$ for $s = (-1, -0.9]$. Note that we used the actual frequency regulation signal to simulate the VPP's response, which may be biased from the predicted distribution (Fig.~\ref{fig_signal}). To cope with this, the VPP updates its bids for future periods every half hour based on the latest resource states.

\subsection{Results and Comparisons}

\begin{small}
  \begin{table}[!t]\renewcommand{\arraystretch}{0.85}
    \caption{Comparison of VPP Profits Using Different Methods}
    \label{tab_profit_method}
    \centering
    \begin{tabular}{cccc}
      \toprule
      \multicolumn{1}{c}{\multirow{2}{*}{\begin{tabular}[c]{@{}c@{}}Bidding-Disaggregation \\      Method\end{tabular}}} &
      \multicolumn{1}{c}{\begin{tabular}[c]{@{}c@{}}Income from\\      the Market\end{tabular}}                  &
      \multicolumn{1}{c}{\begin{tabular}[c]{@{}c@{}}Operational\\      Cost\end{tabular}}                  &
      \multicolumn{1}{c}{\begin{tabular}[c]{@{}c@{}}VPP \\      Profit\end{tabular}}                                                                                                   \\
      \multicolumn{1}{c}{}                                            & \multicolumn{1}{c}{(\$)} & \multicolumn{1}{c}{(\$)} & \multicolumn{1}{c}{(\$)} \\ \midrule
      \multirow{2}{*}{Proportional Disaggregation}                    & {1709}                   & {1455}                   & {254}                    \\
                                                                      & (0.0\%)                  & (0.0\%)                  & (0.0)                    \\ \cline{2-4}
      \multirow{2}{*}{Greedy Disaggregation}                          & 1316                     & 1845                     & -529                     \\
                                                                      & (-23.0\%)                & (+26.8\%)                & (-782)                   \\\cline{2-4}
      \multirow{2}{*}{\textbf{Optimal Disaggregation}}                & \textbf{2622}            & \textbf{1289}            & \textbf{1333}            \\
                                                                      & \textbf{(+53.3\%)}       & \textbf{(-11.4\%)}       & \textbf{(+1079)}         \\ \bottomrule
    \end{tabular}
  \end{table}
\end{small}

\begin{figure}[!t]
  \centering
  \subfloat[]{
    \includegraphics[width=2.5in]{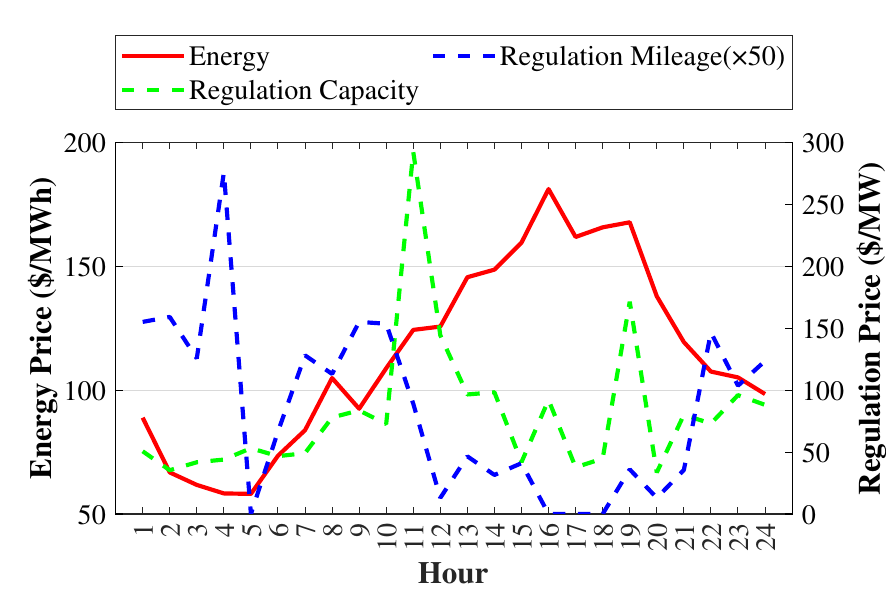}}
  \subfloat[]{
    \includegraphics[width=2.5in]{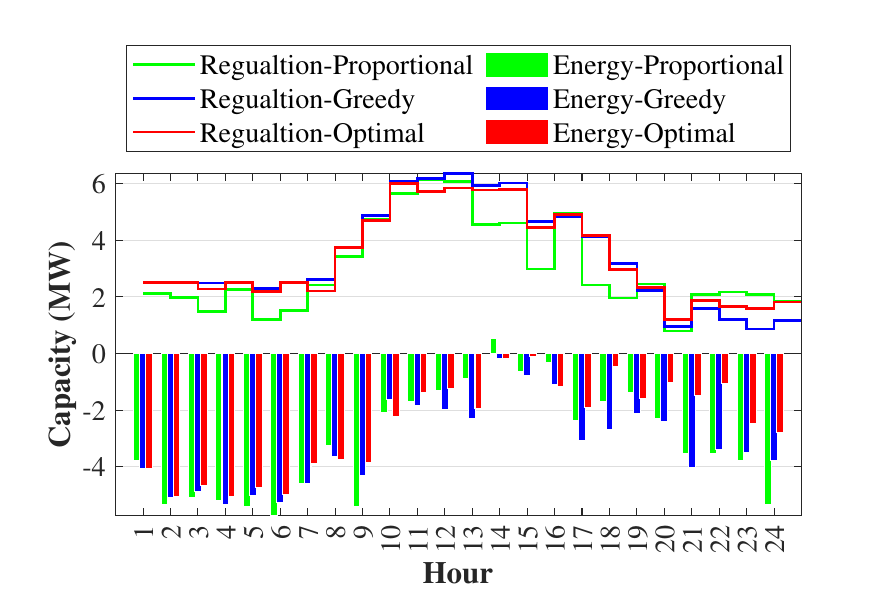}}\newline
  \subfloat[]{
    \includegraphics[width=2.5in]{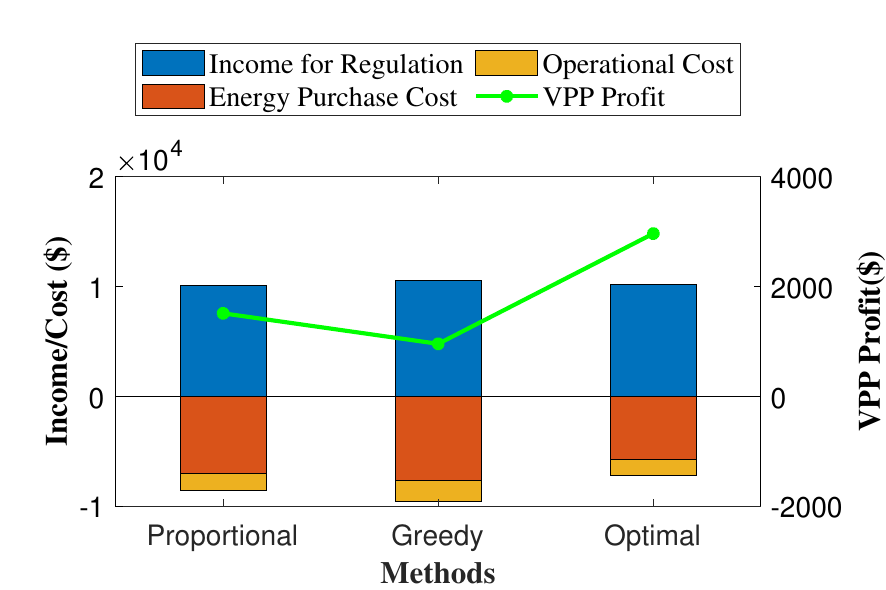}}
  \subfloat[]{
    \includegraphics[width=2.5in]{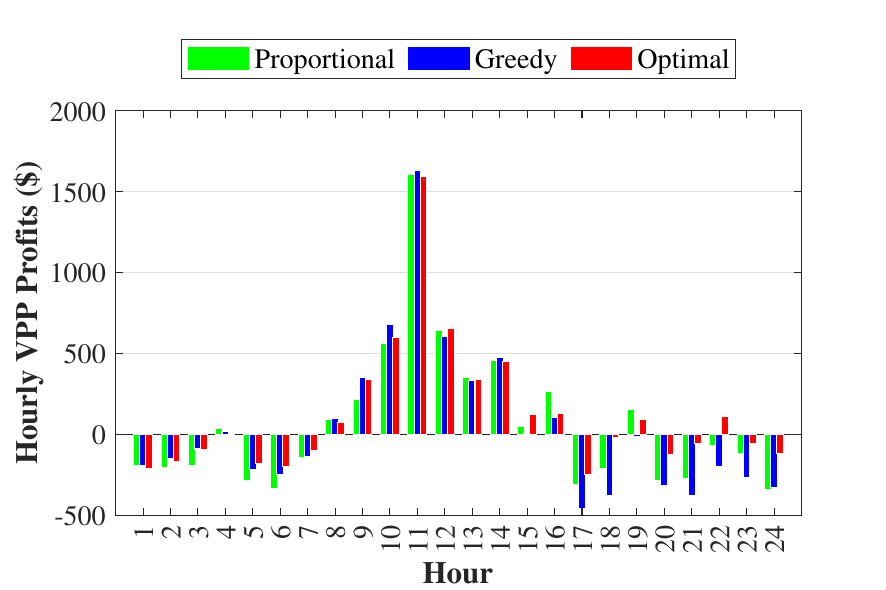}}
  \caption{Results of July 21: (a) Real-time energy and regulation market prices. (b) Final bids of the VPP in the electricity market. (c) Market income and operational costs throughout the day. (d) Hourly profits of the VPP.}
  \label{fig_price_bid}
\end{figure}

Under the above settings, we run three different strategies for providing regulation services, calculate the costs and revenues of the VPP, and summarise the average results over July 15 - July 28 in Table~\ref{tab_profit_method}, with the bidding results presented in Fig.~\ref{fig_price_bid}(b). Here, \textbf{Optimal Disaggregation} refers to the proposed method, i.e., using (\ref{model_bidding_opdis}) to decide the VPP's bids and running Algorithm~\ref{alg_control} to respond to the frequency regulation signal. The difference between \textbf{Greedy Disaggregation} and Optimal Disaggregation is that $\lambda^e_{t, j}$ in (\ref{model_substitution}) is set to 0, so the strategy aims to minimise the immediate response cost. \textbf{Proportional Disaggregation} is a commonly used strategy in the existing literature~\citep{yi_aggregate_2021}.

It can be seen from Table~\ref{tab_profit_method} that under optimal disaggregation, the market income of the VPP increases significantly, including income from the regulation market and electricity purchasing expenses. Furthermore, the operational costs also decrease, contributing to a significant improvement in VPP profits compared to those of other methods. Interestingly, using greedy disaggregation, i.e., merely minimising the immediately induced operational costs, leads to the lowest VPP profits, which highlights the necessity of considering the temporal-coupling characteristics of VPP operation. In contrast, optimal disaggregation enables maximisation of the VPP profit throughout the day (Fig.~\ref{fig_price_bid}(c)), although its profits may not be the highest in some time periods (Fig.~\ref{fig_price_bid}(d)). 

\subsection{Regulation Deployment Results}

\begin{figure}[!t]
  \centering
  \subfloat[]{
    \includegraphics[width=2.5in]{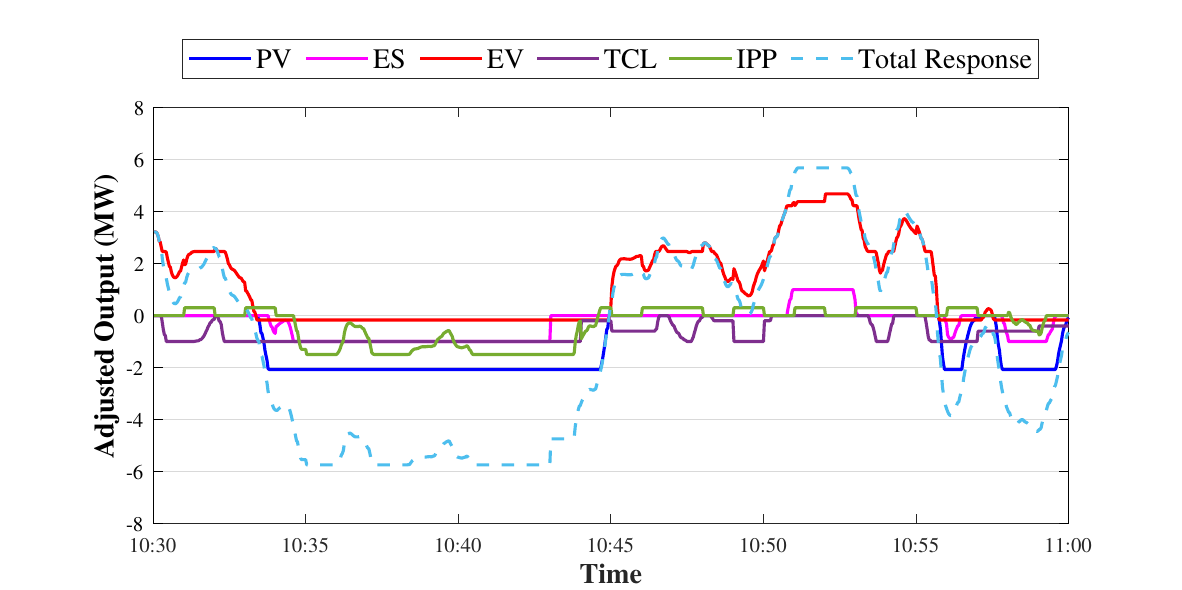}}
  \subfloat[]{
    \includegraphics[width=2.5in]{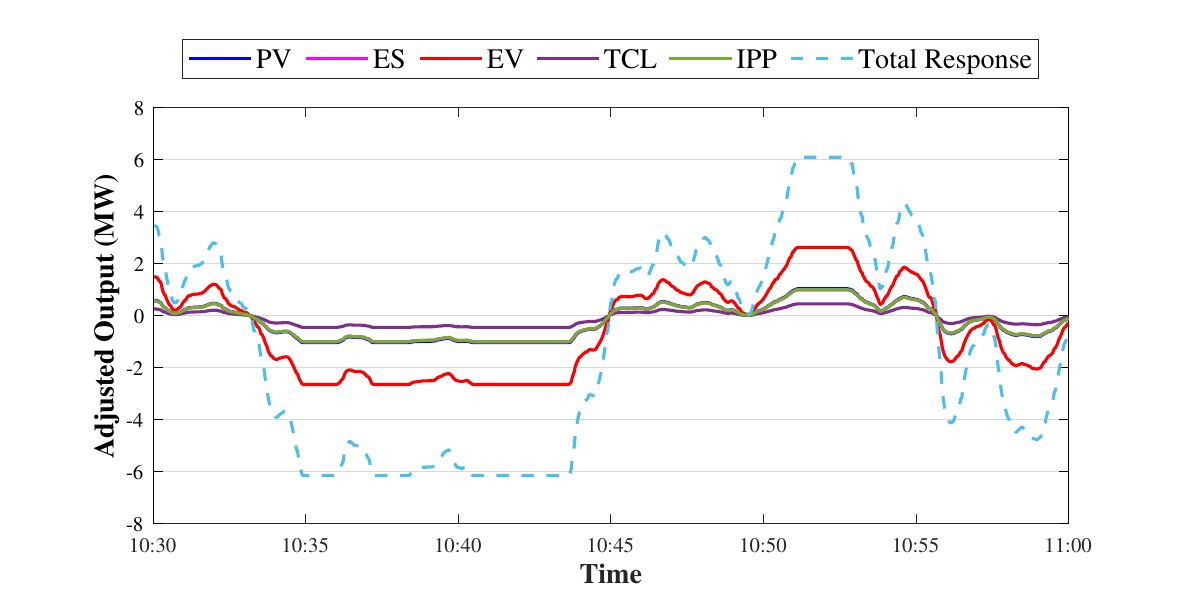}}
  \caption{Power aggregation results for deploying regulation at 10 a.m., July 21: (a) Optimal disaggregation. (b) Proportional disaggregation.}
  \label{fig_typical_hour}
\end{figure}

Fig.~\ref{fig_typical_hour} shows the disaggregation results for 10:30-11:00 on July 21. The optimal disaggregation strategy (a) prioritises the use of low-cost resources. For example, after receiving an up-regulation (injecting energy into the grid) command, EVs respond first, followed by ES. Note that the power adjustment here is relative to the benchmark power; therefore, the \textit{increase} in EV power does not necessarily imply discharging: it may be the reduced charging power, which does not incur additional costs. On the other hand, under proportional disaggregation (b), all resources respond to regulation signals simultaneously, regardless of their cost characteristics.

\begin{figure}[!t]
  \centering
  \subfloat[]{
    \includegraphics[width=1.6in]{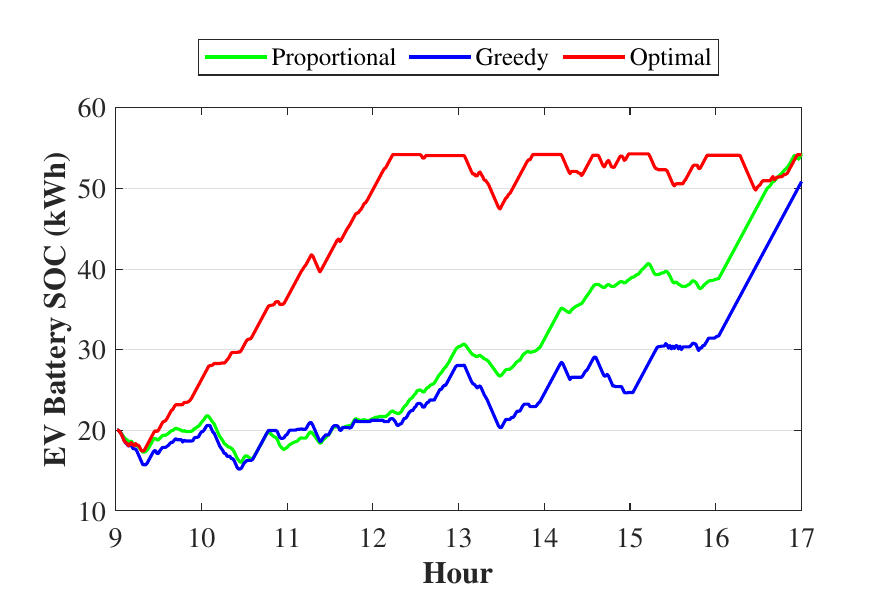}}
  \subfloat[]{
    \includegraphics[width=1.6in]{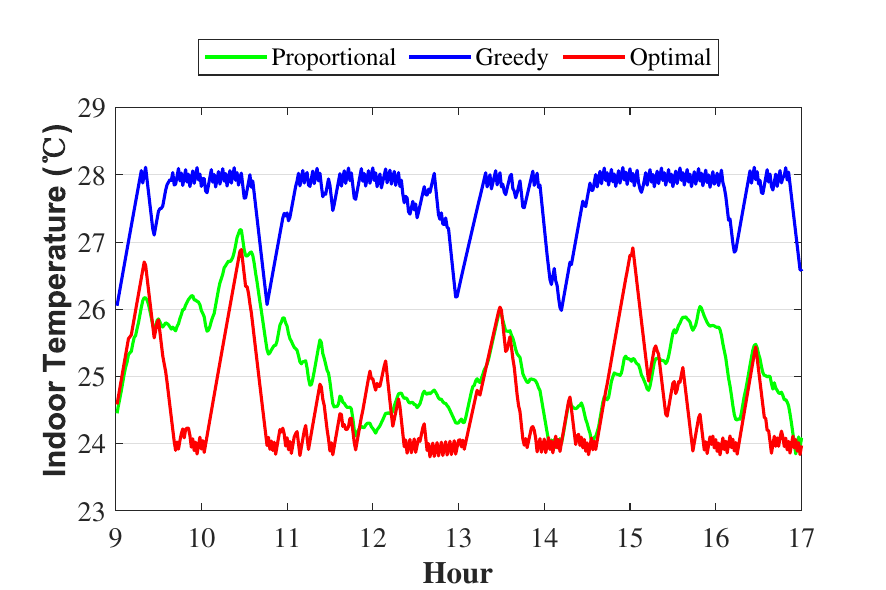}}
  \subfloat[]{
    \includegraphics[width=1.6in]{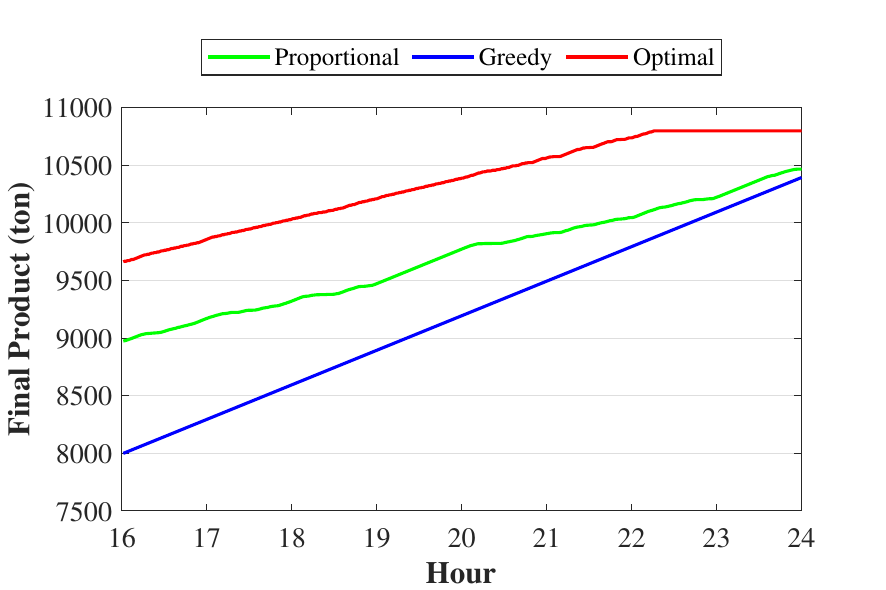}}
  \caption{Change in resource states when deploying regulation on July 21: (a) State of charge of an EV. (b) Indoor temperature of a TCL. (c) Final product of the IPPs.}
  \label{fig_state}
\end{figure}

Fig.~\ref{fig_state} illustrates the influence of regulation deployment on the SOC of EV batteries and the indoor temperature of TCL under different disaggregation strategies. From this, we can intuitively understand the advantages of optimal disaggregation considering the temporal-coupling characteristics of resources, especially compared to greedy disaggregation. In (a), under greedy decomposition (blue), the SOC of the EV battery remains at a low level at 16:00, which is caused by frequent reduction of EV charging power. To achieve the target SOC before departure, the EV must maintain maximum charging power until 17:00 and is thus unable to provide regulation capacity. Similarly, greedy disaggregation (b) reduces TCL power at 10:00, causing the indoor temperature to reach the set threshold by 11:00, affecting its regulation capacity at that time. The results of IPPs (c) are similar, where the last IPP under greedy disaggregation must operate at rated power in the last few hours to achieve production targets. This can explain the higher VPP profits observed in the last few hours of the day under the optimal disaggregation, as shown in Fig.~\ref{fig_price_bid}(c).

\subsection{Sensitivity Analysis}
The fluctuation in the degradation price of the ES can impact the profitability of the VPP. We conducted an investigation to assess the influence of these parameters on the performance of our proposed methodology. Fig.~\ref{fig_calculation_time}(a) illustrates the VPP's market income and operational costs corresponding to different degradation prices of the ES, ranging from 0.25 to 4 times the default value.
As the degradation price escalates, the market income diminishes across all methodologies. However, it is noteworthy that our proposed approach consistently outperforms the alternative methods. These outcomes strongly imply the superiority of our approach over traditional methods, which do not incorporate real-time optimisation of power allocation within the VPP's internal resources. 

\subsection{Calculation Time}

\begin{figure}[!t]
  \centering
  \subfloat[]{
    \includegraphics[width=2.5in]{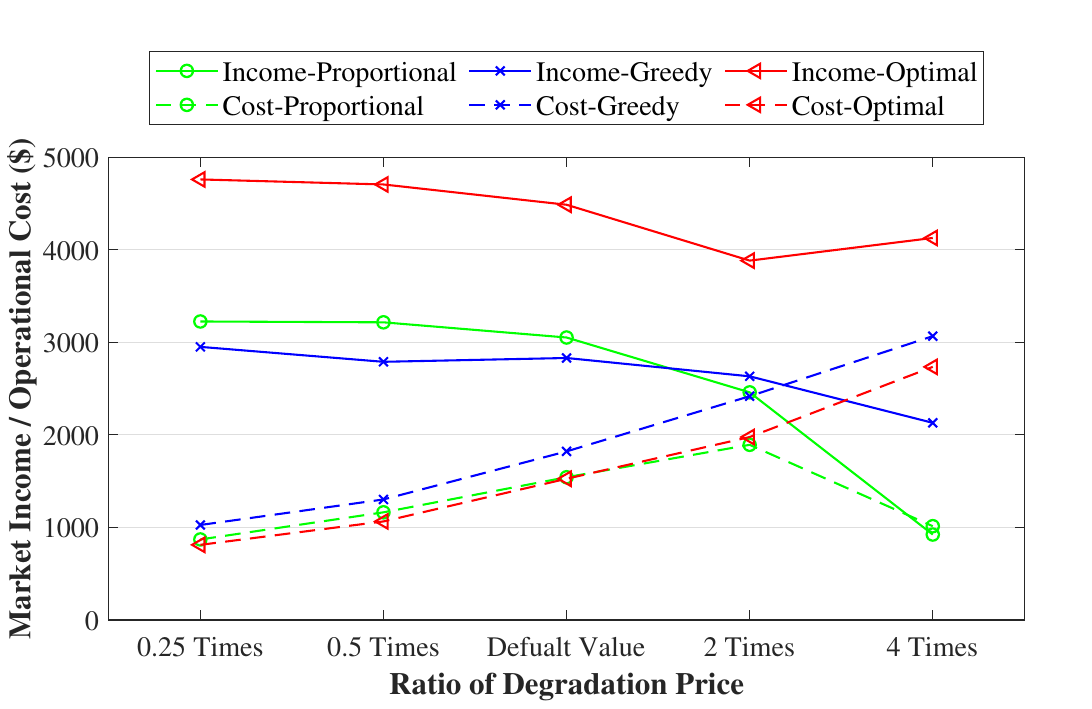}}
  \subfloat[]{
    \includegraphics[width=2.5in]{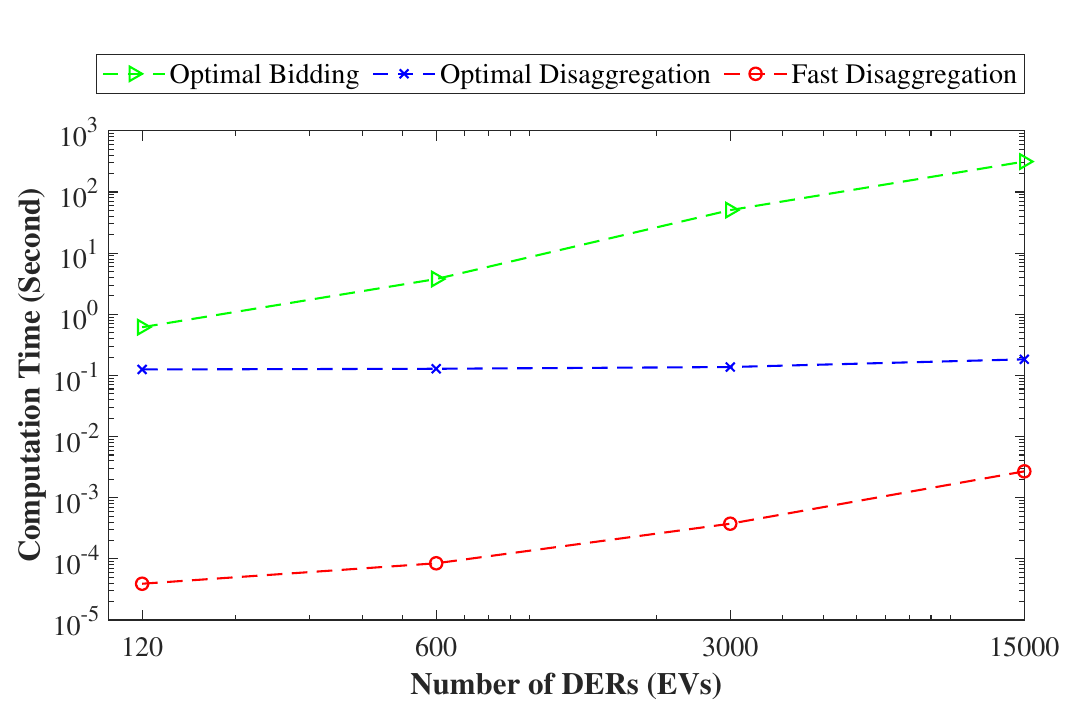}}
  \caption{(a) VPP income from the market and operational costs with different degradation prices. (b)Computation time of the proposed method with increasing number of resources.}
  \label{fig_calculation_time}
\end{figure}

We used CPLEX (V12.4) and MATLAB (R2021a) with YALMIP to solve the optimisation problems on a workstation with an Intel Core i9-10900X CPU (3.7 GHz) and 128 GB RAM.
Fig.~\ref{fig_calculation_time}(b) shows the variation in computation time for solving the optimal bidding problem, optimal disaggregation problem, and fast disaggregation algorithm with the number of resources in the VPP. Without loss of generality, we increased the number of EVs to increase the number of resources since EVs are typical resources with small individual capacities but large quantities.
In the test, the computation time for the original optimal bidding problem exceeded 5 minutes after reaching 15000 EVs, which is unacceptable for online response to frequency regulation signals. However, it is still satisfactory for determining the bids. The optimal disaggregation problem reduced the computation time to less than 1 second but relies on powerful solvers and computing platforms. The fast disaggregation algorithm, on the other hand, does not rely on these resources and further reduces the computation time by several orders of magnitude. The results imply that our method is in line with the response time requirements for providing regulation services.

\section{Conclusion}\label{sec_conclusion}
Throughout the process of delivering real-time services such as frequency regulation, a VPP can harness the heterogeneous operational characteristics of its resources to optimise power disaggregation when responding to signals issued by the GO. This optimisation, in turn, leads to enhanced VPP profitability. In this study, we present an optimal operational strategy for VPPs to provide regulation services, which achieves an optimal disaggregation approach by considering temporal-coupling characteristics.

To meet the stringent response time requirements for regulation deployment, we transform the original stochastic optimisation problem into a smaller-scale linear programming problem of equivalent nature. Furthermore, we employ a rapid disaggregation algorithm that obviates the need for online optimisation solvers. This algorithm ensures efficient real-time solving without compromising computational speed. Notably, our numerical results demonstrate that our method yields substantial improvements in VPP profits while incurring minimal online computational costs.

By replacing the traditional proportional disaggregation paradigm, our approach has the potential to revolutionise the operational strategy of VPPs. This paradigm shift allows VPPs to fully leverage their unique advantages in effectively utilizing diverse resource types in a complementary manner, in stark contrast to conventional power plants.

Future research endeavors should focus on incorporating practical constraints, such as resource ramping rate limitations, and exploring VPP profit allocation strategies within the proposed framework. Such investigations will further enhance the applicability and efficacy of our approach in real-world scenarios.


\section*{Acknowledgment}
This work was supported by the National Key R\&D Program of China (No. 2021YFB2401200) and National Natural and Science Foundation of China (No. 52107102).
Qixin Chen, Ruike Lyu, Hongye Guo, and Xiangbo Su are with
the State Key Laboratory of Power Systems, Department of Electrical
Engineering, Tsinghua University, Beijing 100084, China. Generative artificial intelligence was used to enhance grammar and readability.

\appendix

\section{Compact Formulation of the Standardised Operation Model}\label{app_cptDOFR}
To facilitate readers' comprehension, we presented the standardised operational model for the VPP in an element-wise form in Section\ref{sec_model}. Here we provide a compact matrix-form model (\ref{model_cpt}). The meaning of the model is the same as before and will not be explained here. In the following model, $P^{\rm dis(ch)}_{t} \in R^{S \times I}$, i.e., $[p^{\rm dis(ch)}_{t,s,i}]$ denotes the power matrix for the regulation signal scenarios and the resources at time $t$. Similarly, $E_{t} \in R^{I}$ denotes the state vector of the resources at time $t$.
$H^{\rm dis} \in R^{I \times I}$($[\frac{1}{\eta^{\rm dis}_{ij}}]$) and $H^{\rm ch} \in R^{I \times I}$($[\eta^{\rm ch}_{ij}]$) are the incidence matrices reflecting the power-state transition relationship between the $I$ resources, while $\Theta$ is the state-state transition matrix. Underlined/overlined are the corresponding parameters for lower/upper limits. $\textbf{1}_{\rm I} \in R^{I}$ denotes a vector of all ones. $\delta = [\delta_s]$ is the vector of regulation signals values, and $\Pi_t$ is the vector whose elements are the probability of the corresponding regulation signal scenario at time $t$. $\Theta' = (\frac{\Delta t - \Delta t^{\rm req}}{\Delta t}\textbf{1}_{\rm I} - \frac{\Delta t^{\rm req}}{\Delta t}\Theta)$. $w_t = [w_{t, i}]$. $P^{\rm dis(ch)}_{t,(1/-1)}$ denotes the vector from the power matrix for the regulation signal scenarios of $\delta_s = 1/-1$.
\begin{subequations}\label{model_cpt}
  \begin{equation}\label{model_cpt_powerLimit}
    \underline{P}^{\rm dis(ch)}_{t} \le P^{\rm dis(ch)}_{t} \le \overline{P}^{\rm dis(ch)}_{t}, \ \forall  t
  \end{equation}
  \begin{equation}\label{model_cpt_energyLimit}
    \underline{E}_{t} \le E_{t} \le \overline{E}_{t},\ \forall t
  \end{equation}
  \begin{equation}\label{model_cpt_energyChange}
    E_{t} = \Theta E_{t - 1} +
    (H^{\rm ch} P^{\rm ch}_{t} -  H^{\rm dis} P^{\rm dis}_{t}) \Pi_{t} \Delta t, \ \forall t
  \end{equation}
  \begin{equation}\label{model_cpt_balance}
    (P^{\rm ch}_{t} - P^{\rm dis}_{t})^{\top} \textbf{1}_{\rm I} = \tilde{p}_t \textbf{1}_{\rm S} + r_t\delta, \forall t
  \end{equation}
  \begin{equation}\label{model_cpt_req1}
    \Theta' E_t - H^{\rm dis}P^{\rm dis}_{t,(1)} \Delta t^{\rm req} + w_t \Delta t^{\rm req} \ge \underline{E}_{t}, \forall t
  \end{equation}
  \begin{equation}\label{model_cpt_req2}
    \Theta' E_t + H^{\rm ch}P^{\rm ch}_{t,(-1)} \Delta t^{\rm req} + w_t \Delta t^{\rm req} \le \overline{E}_{t}, \forall t
  \end{equation}
  \begin{equation}\label{model_cpt_cost}
    cost_{t} = \textbf{1}_{\rm I}^{\top} (Pr^{\rm dis}_{i} P^{\rm dis}_{t} + Pr^{\rm ch}_{i} P^{\rm ch}_{t}) \Pi_t \Delta t, \ \forall t
  \end{equation}
\end{subequations}

\section{Proof of Proposition~\ref{theorem_optimality}}\label{theorem_optimality_app}

The idea is to verify that the optimality conditions of the bidding problem can imply the optimality conditions of the optimal disaggregation problem. For the sake of expression, we use (\ref{model_std_balance2}) to eliminate $p_{t,i,s}$. We follow the convention to write the objective function as min. and the inequalities as $\le 0$, despite their expression in the aforementioned format. First, we write out the parts where $\{p^{\rm ch(dis)}_{t,i,s} \ (t=1)\}$ appear in the KKT conditions of the bidding problem:

a) Primal feasibility: (\ref{model_std_powerLimit}) and (\ref{model_std_balance}). Denote the dual variable of the two constraints as $\underline{\mu^{\rm dis(ch)}_{t, s, i}} / \overline{\mu^{\rm dis(ch)}_{t, s, i}}$ and $\lambda^{\rm bal}_{t,s}$, respectively.

b) Dual feasibility: $\underline{\mu^{\rm dis(ch)}_{t, s, i}} \ge 0, \overline{\mu^{\rm dis(ch)}_{t, s, i}} \ge 0, \ \forall i$.

c) Complementary slackness: $\underline{\mu^{\rm dis(ch)}_{t, s, i}} (\underline{p}^{\rm dis(ch)}_{t, i}- p^{\rm dis(ch)}_{t, s, i}) = 0$, $\overline{\mu^{\rm dis(ch)}_{t, s, i}} (p^{\rm dis(ch)}_{t, s, i} - \overline{p}^{\rm dis(ch)}_{t, i}) = 0, \ \forall i$

d) Stationarity: ${\partial L}/{\partial p^{\rm dis(ch)}_{t, s, i}} = 0, \ \forall i$, where $L$ is the Lagrangian function of the bidding problem. We have:
$$\lambda^{\rm bal}_{t,s} - \underline{\mu^{\rm ch}_{t, s, i}} + \overline{\mu^{\rm ch}_{t, s, i}} +  \pi_{t, s}  (Pr^{\rm ch}_{i} + \sum_{j} \lambda^{\rm e}_{t, j} \eta^{\rm ch}_{ij} )\Delta t = 0, \forall i$$
$$- \lambda^{\rm bal}_{t,s} - \underline{\mu^{\rm dis}_{t, s, i}} + \overline{\mu^{\rm dis}_{t, s, i}} +  \pi_{t, s}  (Pr^{\rm dis}_{i} - \sum_{j} \lambda^{\rm e}_{t, j} \frac{1}{\eta^{\rm dis}_{ij}} )\Delta t = 0, \forall i$$

The KKT conditions of the optimal disaggregation problem are as follows:

a) Primal feasibility: (\ref{model_opdis_powerLimit}) and (\ref{model_opdis_balance}). Denote the dual variable as $\underline{\mu^{\rm dis(ch)}_{t, i}} / \overline{\mu^{\rm dis(ch)}_{t, i}}$ and $\lambda^{\rm bal}_{t, s}$, respectively.

b) Dual feasibility and c) Complementary slackness are in the same form as in the bidding problem.

d) Stationarity: ${\partial L}/{\partial p^{\rm dis(ch)}_{t, i}} = 0, \ \forall i$. We have:
$$\lambda^{\rm bal}_{t, s} - \underline{\mu^{\rm ch}_{t, i}} + \overline{\mu^{\rm ch}_{t, i}} + (Pr^{\rm ch}_{i} + \sum_{j} \lambda^{\rm e}_{t, j} \eta^{\rm ch}_{ij} )\Delta \hat{t} = 0, \forall i$$
$$- \lambda^{\rm bal}_{t, s} - \underline{\mu^{\rm dis}_{t, i}} + \overline{\mu^{\rm dis}_{t, i}} + (Pr^{\rm dis}_{i} - \sum_{j} \lambda^{\rm e}_{t, j} \frac{1}{\eta^{\rm dis}_{ij}} )\Delta \hat{t} = 0, \forall i$$

At the optimum of the strictly feasible (Assumption~\ref{assumption_feasible}) bidding problem, $\exists$ $p^{\rm ch(dis)*}_{t,i,s}$, $\lambda^{\rm e*}_{t, i}$, $\underline{\mu^{\rm dis(ch)*}_{t, s, i}}$, $\overline{\mu^{\rm dis(ch)*}_{t, s, i}}$, and $\lambda^{\rm bal*}_{t, s}$, such that the above KKT conditions are satisfied. Then, we can verify that the KKT conditions of the optimal disaggregation problem are also satisfied by setting $p^{\rm ch(dis)}_{t,i,s} = p^{\rm ch(dis)*}_{t,i,s}$, and $(\underline{\mu^{\rm dis(ch)}_{t, s, i}}, \overline{\mu^{\rm dis(ch)}_{t, s, i}}, \lambda^{\rm bal}_{t, s})=
  \Delta \hat{t} / (\Delta t^{\rm cur} \pi_{t^{\rm cur}, s}) (\underline{\mu^{\rm dis(ch)*}_{t, s, i}}, \overline{\mu^{\rm dis(ch)*}_{t, s, i}}, \lambda^{\rm bal*}_{t, s})$. Since the two problems are convex, the optimal value is unique, and the proposition is proved.

\section{Proof of Proposition~\ref{theorem_invariance}}\label{theorem_invariance_app}

The idea is to verify that the optimal solution to the bidding problem Bid($\hat{t}$) also satisfies the optimality conditions for the bidding problem Bid($\hat{t}'$), where only the differences in KKT conditions regarding (\ref{model_std_energyChange}), (\ref{model_std_energyInit}) and the objective function for the current time period need to be investigated.

Denote the optimal solution for Bid(t) as $\{p^{\rm dis(ch)*}_{t, s, i}\}$, $\{\tilde{p}^{*}_{t}, r^{*}_{t}\}$, and $\{e^{*}_{t, i}\}$. From $\hat{t}$ to $\hat{t}'$, due to responding to regulation signals, the states of the resources change by:
\begin{small}
  \begin{align}\label{app_energyChange_1}
    \Delta e_{i} =
    \underset{s}{\sum}\pi_{t, s} \underset{j}{\sum} (\eta^{\rm ch}_{ij} p^{\rm ch*}_{t, s, j} -  \frac{1}{\eta^{\rm dis}_{ij}} p^{\rm dis*}_{t, s, j}) \Delta t', \ \forall i
  \end{align}
\end{small}
where $\Delta t'$,  the remaining time of the current period, is given by $\Delta t - (\hat{t}' - \hat{t})$. After substituting in $\{p^{\rm dis(ch)*}_{t, s, i}\}$, $\Delta t'$, and (\ref{model_std_energyInit}), constraint (\ref{model_std_energyChange}) at $\hat{t}'$ is replaced by:
\begin{small}
  \begin{align}\label{app_energyChange_2}
    e_{t+1, i} = & \theta_i e^{init}_{i} +
    \underset{s}{\sum}\pi_{t, s} \underset{j}{\sum} (\eta^{\rm ch}_{ij} p^{\rm ch*}_{t, s, j} -  \frac{1}{\eta^{\rm dis}_{ij}} p^{\rm dis*}_{t, s, j}) \Delta t \\
    \nonumber    & + w_i \omega_t \ : \ \lambda^{e}_{t, i}, \ \forall t, \forall i
  \end{align}
\end{small}
which is identical to (\ref{model_std_energyChange}) at $\hat{t}$ and therefore satisfied. The same transformation can be applied to the relevant terms in the target function. Therefore, the primal constraints of Bid($\hat{t}'$) are satisfied.

For the KKT conditions with dual variables, take those related to $\{p^{\rm ch(dis)}_{t,i,s}\}$ as an example. As presented in the proof of Proposition~\ref{theorem_optimality}, in the KKT conditions (a-d), the differences in Bid($\hat{t}$) and Bid($\hat{t}'$) lie only in the parameter $\Delta \hat{t}'$. The ratio $\Delta \hat{t}' / \Delta \hat{t}$ is multiplied by the dual variables of Bid($\hat{t}$), and these dual variables satisfy the KKT conditions of Bid($\hat{t}'$). Similarly, other dual constraints of Bid($\hat{t}'$) can be satisfied. Therefore, the optimal solution to Bid($\hat{t}$) also satisfies the optimality conditions for Bid($\hat{t}'$).


\bibliographystyle{elsarticle-num-names}
\bibliography{reference}

\end{document}